\documentclass{kluwer}
\begin{document}
\newcommand{\msun}{M$_{\odot}$}
\newcommand{\sfr}{M$_{\odot}$ yr$^{-1}$}
\def\ltsima{$\; \buildrel < \over \sim \;$}
\def\simlt{\lower.5ex\hbox{\ltsima}}
\def\gtsima{$\; \buildrel > \over \sim \;$}
\def\simgt{\lower.5ex\hbox{\gtsima}}
\begin{article}
\begin{opening}
\title{The Properties of High Redshift Galaxies}            

\author{D. \surname{Calzetti}\email{calzetti@stsci.edu}}
\author{M. \surname{Giavalisco}\email{mauro@stsci.edu}} 
\institute{Space Telescope Science Institute, 3700 San Martin Dr., Baltimore, 
MD 21218, U.S.A.}                               

%\date: rather not

\runningtitle{The High-z Galaxies}
\runningauthor{Calzetti \& Giavalisco}

\begin{abstract} 
In recent years, a variety of techniques at optical, near-infrared,
sub-mm, and radio wavelengths have opened complementary windows on the
high-redshift Universe. Here we review the current understanding of
the general properties of the z$\gtrsim$2 galaxies detected in the
optical (Lyman-break galaxies) and in the sub-mm (SCUBA sources). We
list some of the key questions that need to be answered in order to
understand the nature and evolution of the high-redshift
galaxies. Wherever possible, we present tentative answers given so far
to those questions, in particular on the low-redshift couterparts of
the high-redshift galaxies, on the impact of dust obscuration on the
observed quantities, and on physical characteristics of the high-z
systems as inferred from observations.
\end{abstract}


\end{opening}

\def\putplot#1#2#3#4#5#6#7{\begin{centering} \leavevmode
\vbox to#2{\rule{0pt}{#2}}
\includegraphics{#1}
\end{centering}}
% e.g., \putplot{psfile}{vspace}{angle}{hscale}{vscale}{hoffset}{voffset}
% with vspace in any TeX units, angle in degrees, scale in percent,
% and offset in PostScript points (72/in)
%

\section{Introduction}
This writing attempts at summarizing the lively discussion that followed
 the session on ``The Evolution of Galaxies with Redshift'', the last one 
of a very stimulating Conference. Four major areas of discussion were 
identified during the session: 
\begin{enumerate}
\item Identification of the low redshift counterparts of the high 
redshift galaxies;
\item The impact of dust on the interpretation of the observables;
\item The nature of the high redshift galaxies; and
\item The evolution of galaxies and of the intergalactic medium (IGM) with 
redshift.
\end{enumerate}
For each of these topics, we will list some of the extant, unanswered
questions and provide, wherever possible, what it is felt are
preliminary answers. When talking of ``high redshift galaxies'', we
will mainly refer to galaxies at redshift z$\gtrsim$2.

\section{The Low-z Counterparts of High-z Galaxies}

The two major questions in this area are:

$\bullet$ Which are the optimal low redshift templates of the distant 
galaxies and how reliably  have they been determined? 

$\bullet$ Is there a difference between the spectral templates and the 
morphological templates? What are the differences? 

The first question has been addressed by a number of authors, and their 
findings are summarized below. 

The galaxy population identified with the Lyman-break technique at
redshifts z$\sim$3 and z$\sim$4 (Steidel et al. 1996, 1999) has been
likened to local starburst galaxies. The selection technique itself
biases the candidates towards active star-forming objects with
moderate amount of reddening by dust, due to the need of observing a
measureable Lyman discontinuity in the restframe 912~\AA. More
specifically, the distant galaxies resemble the nearby `UV-bright'
starbursts, i.e. those with average A$_V\lesssim$3~mag, in terms of
UV stellar absorption features (Steidel et al. 1996), UV
stellar continuum slope distribution (Meurer et al. 1999, Steidel et
al.  1999), and optical nebular emission lines (Pettini et
al. 1998). For reference, Figure~1 shows the comparison between the UV
spectra of the z$\sim$2.7 Lyman-break galaxy MS~1512-cB58 and the
local starburst dwarf NGC5253 (Tremonti et al. 2000); the similarity
between the two spectra is pretty striking, with the major differences
due to the stronger interstellar absorption lines in the distant
galaxy. 

From a physical point of view, star formation rates (SFRs) per unit
area in the Lyman-break galaxies (LBGs) are of order
a~few~M$_{\odot}$~yr$^{-1}$~kpc$^{-2}$, or $\sim$10\% the maximum SFR
per unit area measured in the local Universe (Lehnert \& Heckman 1996,
Meurer et al. 1997).  This value is also relatively similar to what
measured in the local UV-bright starbursts. Finally, blueshifts in the
UV interstellar absorption lines of MS~1512-cB58 have been interpreted
as bulk gas outflows with velocity v$\sim$200~km~s$^{-1}$ (Pettini et
al. 1998), very similar to what observed in local FIR-bright
starbursts (Heckman et al. 1990). From a purely phenomelogical point
of view, what sets LBGs apart from local starbursts is the physical
extent of the star formation: in nearby objects, starbursts are
generally concentrated within the inner kpc$^2$, confined in the inner
region of solid body rotation (Lehnert \& Heckman 1996) of the host
galaxy; star formation covers areas $\approx$10-15~kpc$^2$ in the
distant galaxies (Giavalisco et al. 1996), and apparenly is extended
to most of the ``visible'' part, as the comparison between the
rest--frame UV and optical light suggests (Dickinson 2000; Giavalisco
et al. in prep.). This translates into dust--corrected global SFRs
that are on average 5--10 times larger in LBGs than in local UV-bright
starbursts, and closer to the values measured in FIR-bright starbursts
(e.g., NGC1614).

\begin{figure}
\putplot{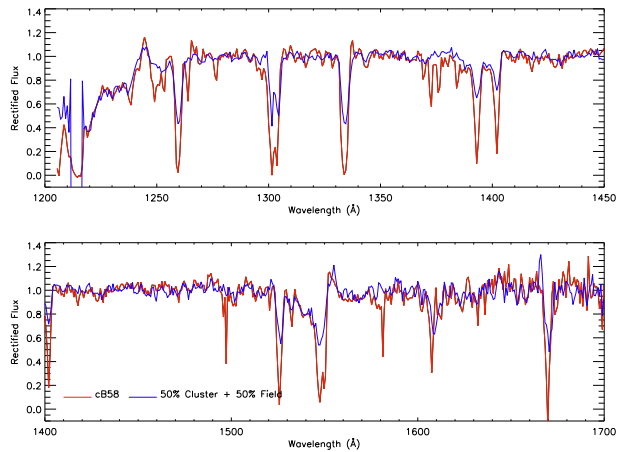}{5.2 cm}{90}{35}{35}{300}{-23}
\caption{The comparison between the UV spectrum of the z=2.7 galaxy
MS~1512-cB58 and that of the local (D$\sim$4~Mpc) starburst dwarf
NGC5253 shows the similarity between the two (Tremonti et
al. 2000). The local starburst UV spectrum has been constructed from a
long-slit HST/STIS spectrum, by requiring that 50\% of the UV light
comes from stellar clusters and 50\% from the diffuse stellar
population, as observed in starbursts (Meurer et al. 1995). The main
difference between the two spectra is in the intensity of the
interstellar absorption lines, that are much stronger in the distant
galaxy.}
\end{figure}

The similarity between local starbursts and LBGs brings
forward another consideration. The number density at the bright end of the
z$\sim$3 galaxies is similar to that of the z$\sim$4 galaxies (Steidel et
al. 1999), and the time interval between z=4 and z=3 is $\sim$350-600~Myr,
depending on the cosmology. Preliminary results suggest that star formation
can last a few 100~Myr in local starbusts (Calzetti 1997, Calzetti et
al. 1997).  If this is true also for the LBGs (Dickinson 2000,
private communication), by z=3 these systems have formed
$\gtrsim$10$^{10}$~M$_{\odot}$, or a large fraction of the stellar mass of an
L$^*$~galaxy.

Increasing evidence suggests that dust-corrected LBGs may account for
$\sim$50\% of the star formation at z$\ge$2, with the other $\sim$50\%
provided by the FIR-bright SCUBA sources (e.g., Barger et al. 2000 and
references therein). Within the uncertainties, actual numbers are in
the 20-80\% range for both type of objects, depending on the adopted
dust correction for the UV-bright galaxies and on the adopted
completeness corrections and AGN fraction for the FIR-bright objects
(Hughes et al. 1998, Almaini et al. 1999, Barger et al. 2000). The
brightest SCUBA sources detected at 850~$\mu$m seem to be
characterized by very faint optical/near~infrared emission and by
thermal spectral energy distributions in the far infrared (Barger et
al. 2000). It has been suggested that their most immediate local
counterparts are the Ultraluminous Infrared Galaxies (ULIRGs, Sanders
\& Mirabel 1996).

\section{The Impact of Dust}

Here the two main questions are:

$\bullet$ How relevant is dust and dust obscuration in the high-z galaxies?

$\bullet$ Is there dust in the IGM, and at what level?

Measurements of the Cosmic Infrared background with COBE (Fixsen et
al. 1998, Hauser et al. 1998) have shown that the amount of energy
detected beyond 40~$\mu$m is comparable or higher (up to a factor of
$\sim$2) than the UV-optical background (Pei, Fall \& Hauser
1999). Thus the stellar energy absorbed by dust and re-radiated in the
FIR represents a non-negligible ingredient in the energy balance of
galaxies at all redshifts.

LBGs are characterized by a distribution of UV stellar continuum
slopes with median value $\beta\sim -$1.4 (Dickinson 1998), much
redder than the value $\beta_o\sim -$2.1 expected for a dust-free star
forming population (Leitherer \& Heckman 1995). In local starburst
galaxies, the measured slope of the UV stellar continuum is a
sensitive tracer of dust reddening and obscuration (Calzetti et
al. 1994, 2000, Meurer, Heckman \& Calzetti 1999). If local starbursts
are accurate representations of the LBGs, dust is probably present in
the latter population, to the level of obscuring about 80\% of the UV
light (Steidel et al. 1999). Despite most of their UV stellar light is
reprocessed by dust into the FIR, LBGs do not seem to be prominent FIR
emitters, with predicted 850~$\mu$m fluxes at the $\sim$1~mJy level or
less (e.g., Calzetti et al. 2000, Chapman et al. 2000); indeed they
are mostly undetected with SCUBA (Chapman et al. 2000). However,
because of their large number density,
n(z=3)$\sim$1.2~10$^{-2}$~Mpc$^{-3}$ (for $\Omega_{\Lambda}$=0.7,
$\Omega_{matter}$=0.3, H$_o$=65~km/s/Mpc, Giavalisco et al. 2000),
LBGs can still provide a non-negligible contribution to the FIR
background. Estimates range from 25\% up to most of the 850~$\mu$m
background flux, depending on assumptions on both observables and
theoretical prescriptions for the FIR SEDs (Adelberger \& Steidel
2000, Dunlop 2000). Direct 850~$\mu$m counts done with SCUBA down to
0.25~$\mu$m reproduce almost entirely (94\% of) the COBE background
(Blain et al. 1999), with the brigthest SCUBA sources, those above
2~mJy, accounting for $\sim$30\% of it. These results strongly
indicate that up to 90\% of the early star formation emission has been
reprocessed by dust into the FIR, and that dust is a widespread
constituent of galaxies at z$>$2.

The amount of dust distributed in the IGM is even less constrained.
Metals have been observed in z=3 Lyman-$\alpha$ Forest clouds down to
column densities N(HI)$\sim$10$^{14.5}$~cm$^{-2}$ (Ellison et
al. 2000). In hierarchical CDM models these clouds are naturally
arising as a consequence of the growth of density fluctuations in the
presence of a UV ionizing background (Hernquist et al. 1996). The
processes for polluting the high-z IGM with metals and, therefore,
dust have not been completely clarified yet. The proposed scenarios go
from widespread metal injection by Pop.~III stars or by subgalactic
structures to in-situ pollution by metal-enriched, supernova-driven
gas outflows from the early galaxies (Gnedin \& Ostriker 1997, Madau
et al. 2000). Whatever the mechanism, the resulting metallicity of the
IGM at z$<$0.5 appears to be $\approx$10\% solar (Barlow \& Tytler
1998) or possibly higher. Much less agreed-upon is the amount of dust
that went into the IGM with the metals (conditions in the IGM do not
favor dust formation, see Aguirre 1999). The evolution of the quasars
UV spectral index indicates that the total IGM dust opacity is
A$_V$(z$<$1.5)$<$0.05, for a Milky~Way-type dust (Cheng et
al. 1991). Aguirre (1999) argued that sputtering in galaxy haloes and
in the IGM could destroy the small grain dust population, thus
changing the extinction curve to a grey one; in particular, small
reddening values, E(B$-$V)$\sim$0.02, would co-exist with large
obscurations, A$_V\sim$0.3. However, a recent analysis of distant Type
1A SNe (Riess et al. 2000) seems to disfavor the substantial IGM dust
optical depth predicted by Aguirre, even in the presence of grey
opacity.

\section{The Nature of the High-z Galaxies}

There are a number of important and unanswered questions in this area, and 
a few are:

$\bullet$ What is the nature of the high-z galaxies?

$\bullet$ Are the UV-bright (Lyman-break) galaxies linked to the FIR-bright 
(SCUBA) ones?

$\bullet$ How accurate are the metallicity measurements in galaxies and IGM, 
and what are these measurements telling us?

$\bullet$ What is the nature of the Damped Lyman-$\alpha$ systems?

$\bullet$ From an observational point of view, how can we access the 
critical redshift range between z=1 and z=2? 

In this contribution we will address only the first two questions.

The mass spectrum of the high redshift galaxies heavily bears on the
first question. Current estimates are very uncertain, and no
conclusions have yet been reached. Inferred masses of relatively
bright LBGs based on the width of the optical nebular lines (Pettini
et al. 1998) suggest values around $10^{10}$--$10^{11}$ \msun, one to
two order of magnitudes smaller than the dynamical mass of local
bright (i.e. $L^*$) galaxies.  However, the reliability of the nebular
lines kinematics as tracer of the galaxy dynamics has not been
demonstrated yet. The strong spatial clustering of the LBGs
(Giavalisco et al. 1998, 2000; Adelberger et al. 1998) suggest values
in the range $10^{11}$--$10^{12}$ \msun for the bright LBGs, depending
on the cosmology (for $\Omega_{matter}\sim 0.3$, $\Omega_{\Lambda}\sim
0.7$, bright LBGs have dynamical masses $\sim 10^{12}$\msun).

Measuring dynamical masses of distant galaxies with traditional
methods requires observing rotation curves with very high angular
resolution and/or velocity dispersions from spectroscopy of stellar
absorption or nebular emission lines. It also requires to detect the
outer regions of the galaxies, which are strongly biased againts by
the $(1+z)^4$ cosmological surface brightness dimming. It is a rather
difficult task for the 8--meter class telescopes.  Very likely it will
be NGST that will enable systematic kinematical observations of
distant galaxies, if a near-infrared spectrograph with spectral
resolution R$\sim$3,000--5,000 and coverage up to $\sim$5~$\mu$m will
be part of the instruments complement (Stiavelli, 1998). With such an
instrument, the kinematical measurements of Milky~Way-like galaxies
will be possible up to z$\sim$5 and of LMC-like galaxies up to
z$\sim$1--2, using the restframe H$\alpha$ emission. Measuring masses
of, e.g. z$\sim $3 LBGs will settle the fundamental, ongoing debate
over the nature of these systems: whether they are the massive
progenitors of today's spheroids (Giavalisco et al. 1996) or the
low-mass fragments of present-day galaxies predicted by the
hierarchical structure formation in CDM models (e.g., Lowenthal et
al. 1997).

The morphology of the LBGs also offers indication as to their
nature. Imaging with {\it HST} and WFPC2 has shown that these galaxies
exhibit a variety of UV morphologies, although only galaxies fainter
than $m^*$\footnote{The luminosity parameter of the Schechter fit to
the UV luminosity function of the z$\sim 3$ LBGs, which is $m^*=24.5$
(Steidel et al. 1999).}  have been observed. Some galaxies have
compact and regular morphology, with light profiles well described by
an $r^{1/4}$ or an exponential law (e.g. Giavalisco et al. 1996).
Other galaxies have fragmented and distorted morphology, sometime with
multiple components embedded in irregular diffuse light, suggestive of
merging and interaction (Lowenthal et al. 1997). The size is, in
general, smaller than today's bright galaxies, with half--light radii
$0.2^{\prime\prime}\simlt r_{1/2}\simlt 0.4^{\prime\prime}$ (at $z=3$,
1$^{\prime\prime}=8.3$~kpc in the cosmology adopted in this
paper). Thus, the regions where stars are being formed in these
objects are larger than local dwarf and irregular galaxies, but
smaller than the Milky Way, although the reported size is probably an
underestimate of the true one, because of the $(1+z)^4$ surface
brightness dimming. Very interestingly, imaging with NICMOS on {\it
HST} has revealed that the rest--frame optical morphology in almost
all of the observed cases is essentially the same as the UV one (see
Figure~2 and Dickinson 2000). This is consistent with these objects
being relatively young galaxies, and suggests that the ongoing
activity of star formation (traced by the UV light) is the one that
assembled most of the galaxies' structures.

\begin{figure}
\putplot{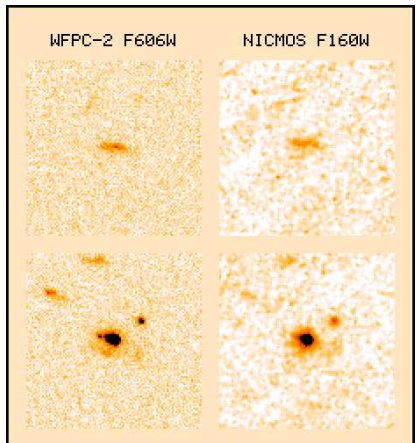}{5.2 cm}{0}{33}{33}{75}{-50}
\caption{A comparison between the rest--frame UV (left) and optical (right)
morphology of Lyman--break galaxies. The shown examples are 0000-2619-C10 at 
$z=3.238$ (top), and 0000-2619-D6 at $z=2.969$ (bottom). Overall, the UV and
optical images are the same, showing that morphological segregation due to a
spatially variable obscuration is not a factor in these galaxies. Dickinson 
(2000) finds the same result from a larger sample from the HDF.}  
\end{figure}

The metallicity is another important parameter to understand the
nature of LBGs and to establish an evolutionary link with present--day
galaxies. Currently the distribution of the metallicity of these
galaxies is essentially unknown, and only a few measures have been
made. Kobulnicky \& Koo (2000), and Teplitz et al. (2000) estimate the
metallicity of the nebular gas from the optical emission lines in a
few objects, finding values in the range $0.3\simlt
Z_{LBG}/Z_{\odot}\simlt 1$. Thus, the gas out of which the last
generation of stars has just formed has already reached a considerable
degree of enrichment. This conclusion is supported by the estimate by
Pettini et al. (2000), who find similar values of metallicity from
stellar and interstellar lines from the UV spectrum of
MS~1512--cB58\footnote{While the agreement is certainly encouraging,
it should taken with caution, because of the possibility that the
interstellar lines used in the analysis are saturated. See the
discussion in Pettini et al. (2000).}. This metallicity is higher than
that found in the local HII galaxies, and is similar to that observed
in Galactic bulge stars and metal--rich globular clusters (e.g. see
the discussion in Kobulnicky \& Koo 2000).

All of the above suggests that the stellar systems observed as LBGs at
$z\sim 3$ would appear today as bright, old and relatively metal--rich
objects. The dust-corrected SFRs, the preliminary indications about
the total mass, the observed morphology, and the fairly large
metallicities are all in qualitative agreement with the interpretation
that these systems have the characteristics of the progenitors of the
spheroids, i.e. bulges of $L^*$ galaxies and ellipticals.

The link between the LBGs (or galaxy fragments) and the SCUBA sources
is also a controversial issue. To the level that the SCUBA sources are
mostly powered by star formation (i.e, the AGN fraction can be
considered small, see Almaini et al. 1999) the two classes of objects
could be considered the two extremes of the starburst phenomenon, with
the UV-bright end occupied by the bluest LBGs and the FIR-bright end
occupied by the most luminous SCUBA sources. Red LBGs and faint SCUBA
sources could lie in between a continuum of properties parametrized by
the SFR and by the dust (metal) content. This scenario is in agreement
with the two recent results that most of the SCUBA sources have very
faint optical/near~infrared counterparts (Barger et al. 2000) and that
the FIR detectability of the Lyman-break galaxies is low (Chapman et
al. 2000). In the local Universe, the positive correlation between
SFR, dust (metal) content, and mass of the host galaxy is a known
property of starbursts (Heckman et al. 1998). Care, however, should be
taken in concluding that the LBGs represent only or mainly low-SFR
(low-mass?) systems; among these galaxies there are many with
dust-corrected SFRs of a few 100~M$_{\odot}$~yr$^{-1}$ (e.g.,
Dickinson 1998), thus comparable in bolometric luminosity to ULIRGs,
although with much less extreme values of the dust obscuration.

While the space density of LBGs is comparable to that of local
galaxies (n$_{local}\sim$1.6$\times$10$^{-2}$~Mpc$^{-3}$ for our
cosmology, Marzke et al. 1994), the space density of the bright
($>$6~mJy) SCUBA sources at z=1--3 is $\approx$100 times higher than
that of local ULIRGs (n$_{SCUBA}\sim$1.1$\times$10$^{-5}$~Mpc$^{-3}$,
see Barger et al. 2000). Although this figure is potentially affected
by large uncertainties due to the small number statistics of the SCUBA
detections, the FIR-bright objects appear to have been more common at
high redshift than at low redshift.  A scenario that can fit both the
UV-bright and the FIR-bright sources within the cosmological context
is one where the LBGs are the progenitors of the present--day bulges
and elliptical galaxies of intermediate to normal luminosity (around
$L^*$), while the bright SCUBA sources are more directly related
to massive ellipticals ($L>L^*$, see also Dunlop 2000).

\section{Galaxy and IGM Evolution}

The number of unaddressed issues in the realm of galaxy and IGM
evolution is too large to be summarized by the limited scope of this
discussion. We pose here, without attempting answers, a non-exhaustive
list of the questions that most directly bear on our understanding of
the evolution of the luminous and non-luminous matter.

$\bullet$ How did the Hubble sequence originate? Which are the high-z
progenitors of the local Hubble types? How does the galaxy mass
spectrum evolve with redshift?

$\bullet$ How does the stellar Initial Mass Function (IMF) evolve with
redshift, if it evolves at all? 

$\bullet$ Do we have a reliable inventory of all the metals in the local 
Universe? How well are they accounted for by chemical evolution models?

\acknowledgements

The authors thank the organizers for the invitation to this
stimulating conference and for partial support of their stay in
Granada. D.C. acknowledge the Space Telescope European Coordinating
Facility and M.G. the European Southern Observatory in Garching
(Munich) for their hospitality during part of the writing of this
contribution.  Travel to the conference was partially supported by the
STScI Director's Discretionary Research Funds.

\end{article}
\end{document}